\documentclass[english,aps,prd,superscriptaddress,nofootinbib]{revtex4}
\usepackage[T1]{fontenc}
\usepackage[latin1]{inputenc}
\usepackage{amssymb}

\usepackage{graphicx}
\usepackage{color}
\usepackage{amsmath}
\usepackage{amsfonts}
\usepackage{babel}
\makeatother
\begin{document}

\title{Is it possible to observationally distinguish adiabatic Quartessence from $\Lambda$CDM?}

\author{L. Amendola}
\affiliation{INAF/Osservatorio Astronomico di Roma, Via Frascati 33,
I-00040 Monte Porzio Catone, RM, -- Italy}
\author{M. Makler}
\affiliation{Centro Brasileiro de Pesquisas F\'{\i}sicas, CEP
22290-180, Rio de Janeiro, RJ, Brasil}
\author{R. R. R. Reis}
\affiliation{Universidade Federal do Rio de Janeiro, Instituto de
F\'{\i}sica, CEP 21941-972, Rio de Janeiro, RJ, Brazil}
\author{I. Waga}
\affiliation{Universidade Federal do Rio de Janeiro, Instituto de
F\'{\i}sica, CEP 21941-972, Rio de Janeiro, RJ, Brazil}

\date{\today {}}

\begin{abstract}
The equation of state (EOS) in quartessence models interpolates
between two stages: $p\simeq 0$ at high energy densities and
$p\approx -\rho$ at small ones. In the quartessence models analyzed
up to now, the EOS is convex, implying increasing adiabatic sound
speed ($c_{s}^{2}$) as the energy density decreases in an expanding
Universe. A non-negligible $c_{s}^{2}$ at recent times is the source
of the matter power spectrum problem that plagued all convex
(non-silent) quartessence models. Viability for these cosmologies is
only possible in the limit of almost perfect mimicry to
$\Lambda$CDM. In this work we investigate if similarity to
$\Lambda$CDM is also required in the class of quartessence models
whose EOS changes concavity as the Universe evolves. We focus our
analysis in the simple case in which the EOS has a step-like shape,
such that at very early times $p\simeq0$, and at late times $p\simeq
const<0$. For this class of models a non-negligible $c_{s}^{2}$ is a
transient phenomenon, and could be relevant only at a more early
epoch. We show that agreement with a large set of cosmological data
requires that the transition between these two asymptotic states
would have occurred at high redshift ($z_t\gtrsim38$). This leads us
to conjecture that the cosmic expansion history of any successful
non-silent quartessence is (practically) identical to the
$\Lambda$CDM one.
\end{abstract}
\maketitle
\section{Introduction} In the current standard
cosmological model, two unknown components govern the dynamics of
the Universe: dark matter (DM), responsible for structure formation,
and dark energy (DE), that drives cosmic acceleration. Recently, an
alternative point of view has started to attract considerable
interest. According to it, DM and DE are simply different
manifestations of a single unifying dark-matter/energy (UDM)
component. Since it is assumed that there is only one dark component
in the Universe, besides ordinary matter, photons and neutrinos, UDM
is also referred to as quartessence \cite{makler03}.

A prototype candidate for such unification is the quartessence
Chaplygin model (QCM) \cite{kamenshchik01}. Although this model is
compatible with the background data \cite{makler03b}, problems
appear when one considers (adiabatic) perturbations. For instance,
the CMB anisotropy is strongly suppressed when compared with the
$\Lambda$CDM model \cite{carturan03}. Further, it was shown that the
matter power spectrum presents oscillations and instabilities that
reduce the parameter space of the model to a region very close to
the $\Lambda$CDM limit \cite{sandvik02}. However, these oscillations
and instabilities in the matter power spectrum and the CMB
constraints can be circumvented by assuming silent perturbations
\cite{reis03b,amendola05}, i.e., intrinsic entropy perturbations
with a specific initial condition ($\delta p=0$). In fact, silent
perturbations solve the matter power spectrum problem for more
generic quartessence \cite{reis05}. Efforts to solve the mater power
spectrum problem have also been put forward in \cite{bento} and
\cite{bilic}. However, we understand that these works are not,
strictly speaking, quartessence. In fact, \cite{bento} introduces
what seems to be a particular splitting of the Chaplygin model. It
is a two component system although only one component is perturbed.
A way to implement silent perturbations is presented in
\cite{bilic}, but they use additional fields that can be interpreted
as new matter components.

In this work we present a possible alternative to solve the above
mentioned problems in the context of the more standard adiabatic
perturbations scenario. We shall discuss a model in which the
quartessence EOS changed its concavity in some instant in the past.
We focus our investigation on models with a step-like shape EOS. We
show that, in order to be in accordance with observations, the EOS
concavity change would have occurred at high redshifts. Similarly to
what happens in the Chaplygin case, observations constrain one of
the parameters of the model to such a low value that, at least at
zero and first orders, the step-like model cannot be observationally
distinguished from the $\Lambda$CDM model.

\section{A new type of quartessence} In the quartessence
models explicitly analyzed up to now, the EOS is convex, i.e., is
such that
\begin{equation}
\frac{d^{2}p}{d\rho^{2}}=\frac{dc_{s}^{2}}{d\rho}<0.\label{convex}
\end{equation}
Stability for adiabatic perturbations and adiabatic sound speed less
than $c$ imply
\begin{equation} 0\leq
c_{s}^{2}\leq1.\label{stab}
\end{equation}
Condition (\ref{stab}) and the fact that $p<0$ immediately implies
the existence of a minimum energy density $\rho_{\min}$, once the
energy conservation equation is used. This is a generic result for
any uncoupled fluid model in which $w=w\left(\rho\right)$. It
implies that the $p=-\rho$ line cannot be crossed and that in any
such a quartessence model the minimum value of the EOS parameter is
$w_{\min}=-1$. The convexity condition (\ref{convex}) implies that
$c_{s\; \max}^{2}$ occurs at $\rho=\rho_{\min}$. This last result is
only a consequence of the convexity of the EOS. In this case, the
epoch of accelerated expansion is also a period of high adiabatic
sound speed, causing the oscillations and suppressions in the power
spectrum. However, this property is not mandatory for quartessence.
Models with concavity changing equations of state may have
$c_{s}^{2}$ negligibly small at $\rho\simeq\rho_{\min}$. As we shall
show, it is possible to build models in which a non-negligible
$c_{s}^{2}$ is a transient phenomenon and relevant only at a very
early epoch, such that only perturbations with relatively large wave
numbers (outside the range of current linear power spectrum
measurements) are affected.

\begin{figure}
 \includegraphics[height= 5.5 cm,width=7.5cm]{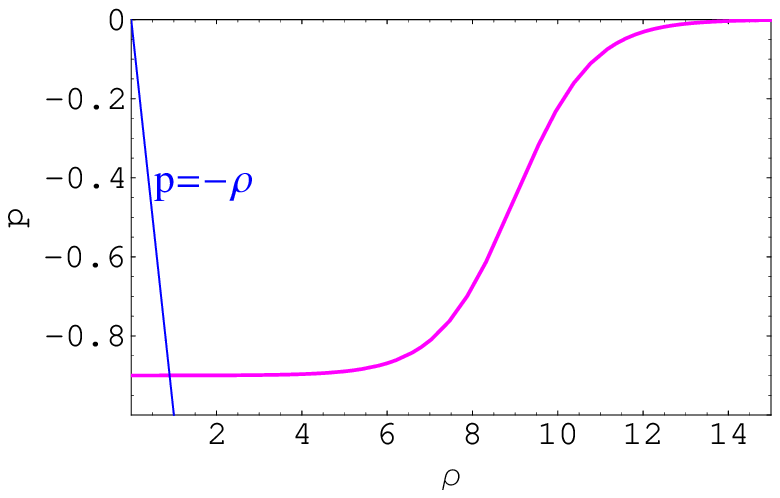} \includegraphics[height=
5.5 cm,width=7.5cm]{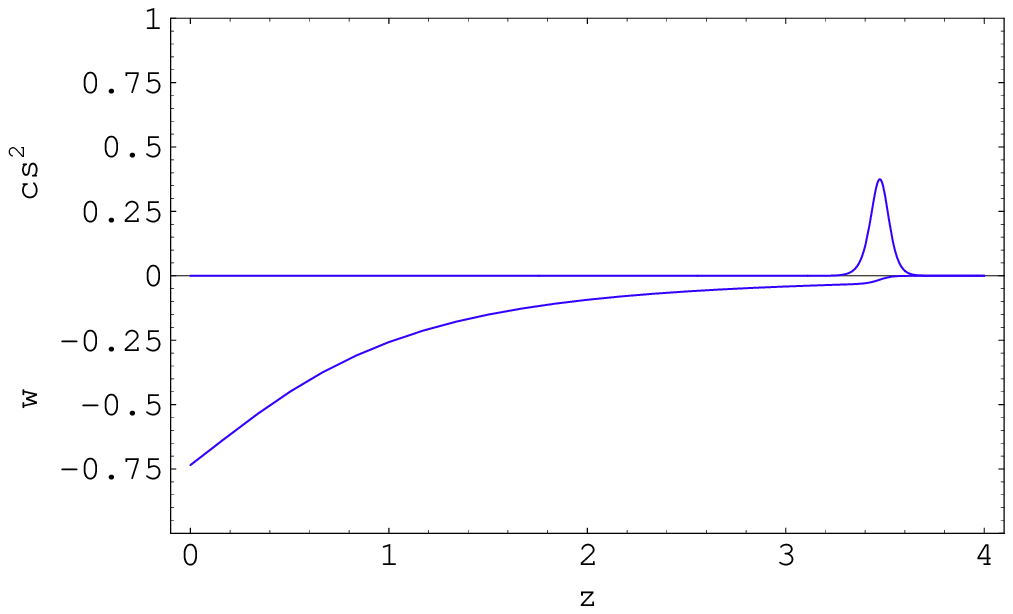} \caption{left panel: pressure ($p$)
as a function of the energy density ($\rho$) for the sigmoid EOS.
Also shown is the $p=-\rho$ line. right panel: typical behavior of
the EOS parameter ($w$) and the adiabatic sound speed ($c_s^2$) as a
function of the redshift ($z$).} \label{pxro}
\end{figure}

The step-like quartessence, given by a sigmoid, is an example of UDM
with concavity changing EOS (see figure \ref{pxro}, left panel),
\begin{equation} p=-M^{4}\left\{
\frac{1}{1+\exp\left[\beta\left(\frac{\rho}{M^{4}}-\frac{1}{\sigma}\right)\right]}\right\}
. \label{sig}
\end{equation}
For this model, the adiabatic sound
speed has the following expression,
\begin{equation}
c_{s}^{2}=\beta\frac{\exp\left[\beta\left(\frac{\rho}{M^{4}}-\frac{1}{\sigma}\right)\right]}
{\left\{ 1+\exp\left[\beta\left(\frac{\rho}{M^{4}}-\frac{1}{\sigma}\right)\right]\right\} ^{2}}.
\label{sigspeed}
\end{equation}
There are three free parameters in the model. The parameter $M$ is
related to the minimum value of the energy density, i.e., the value
of $\rho$ when the asymptotic EOS, $p_{\min}=-\rho_{\min}$, is
reached. The parameter $\sigma$ is related to the value of the
energy density at the transition from the $p\simeq0$ regime to the
$p\simeq-M^{4}$ one ($\rho_{\rm{trans}}=M^{4}/\sigma$ ). Notice that
if $\sigma\ll1$, the transition takes place well before the minimum
density is reached. The parameter $\beta$ controls the maximum sound
velocity $c_{s\max}^{2}$ as well as the redshift width of the
transition region (higher values of $\beta$ implying faster
transitions). For the sigmoid EOS the maximum adiabatic sound speed
is given by $c_{s\; \max}^{2}=\beta/4$, and therefore we require
$0\leq\beta\leq4$.

In the present model, the $\Lambda$CDM limit is not necessarily
associated with the maximum sound speed, in contrast to what is
found in the convex EOS case. The $\Lambda$CDM limit is reached when
$\sigma\rightarrow0$, which implies $p=-\rho=-M^{4}$. Another
possibility is to take $\beta\rightarrow0$. In this case $c_{s\;
\max}^{2}\rightarrow0$ and we also have a $\Lambda$CDM limit, but
now with $p=-\rho=-M^{4}/2$. Since $\beta$ strongly affects the
redshift width of the transition, these two limits have different
characteristics. The case of a nonvanishing $\beta\ll1$ has a
drastic effect on the matter power spectrum. In fact, although the
maximum sound speed will be small, it will be non negligible during
a long redshift range and/or time, practically ruling out these
models.

We note that a step-like quartessence may be represented by the more
generic expression,
\begin{equation}
p=M^{4}f\left[\beta\left(\frac{\rho}{M^{4}}-\frac{1}{\sigma}\right)\right],\label{step}
\end{equation}
where $f$ is a step-like function, with $f(+\infty)=0$ and
$f(-\infty)=-1$. The maximum adiabatic sound speed is $c_{s\;
\max}^{2}=\beta f^{\prime}_{max}$. For $\sigma\ll1$,
$p_{\min}=-M^{4}$.
\begin{figure}
 \includegraphics[height= 6.5 cm,width=8.5cm]{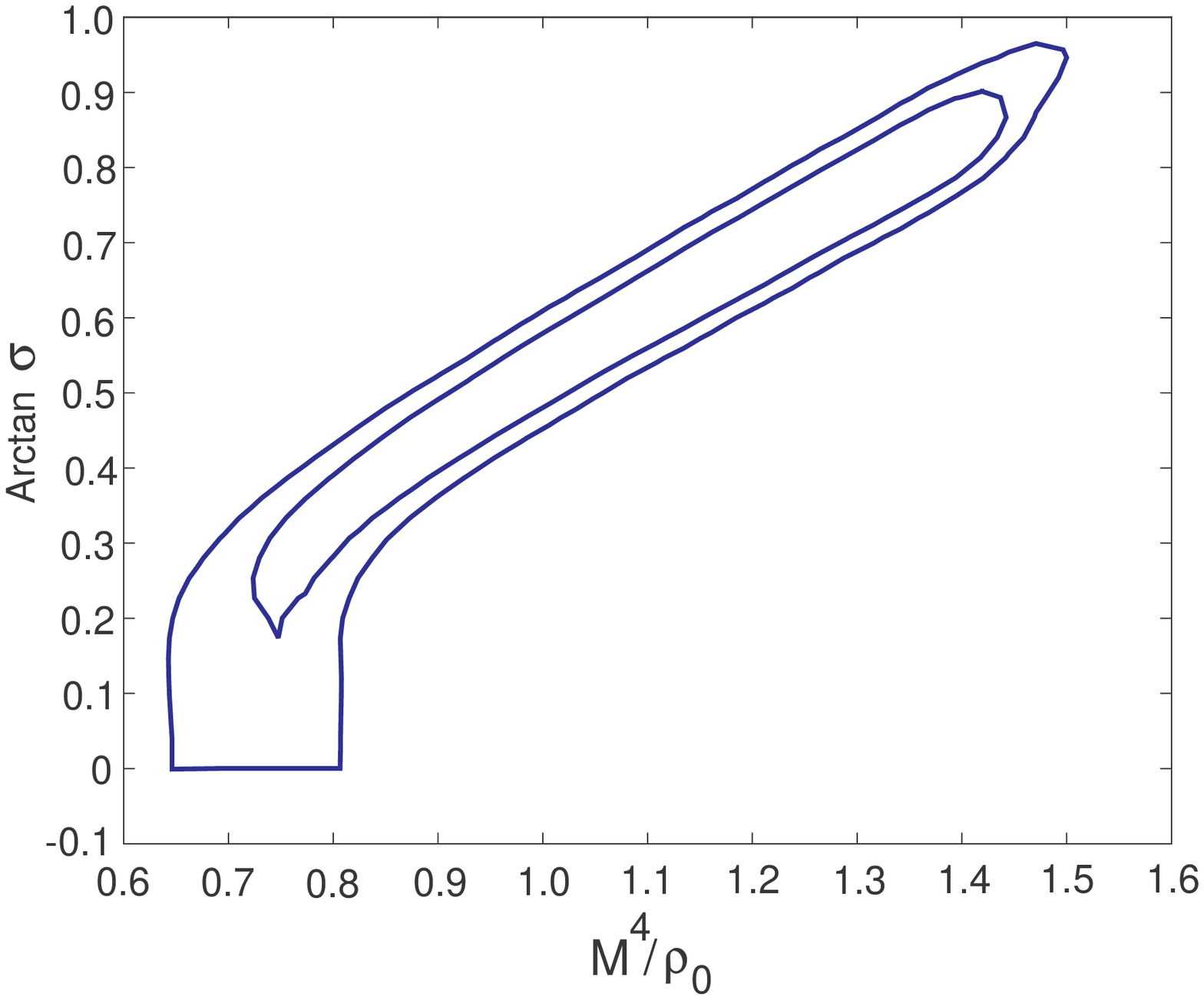} \includegraphics[height= 6.5 cm,width=8.5cm]{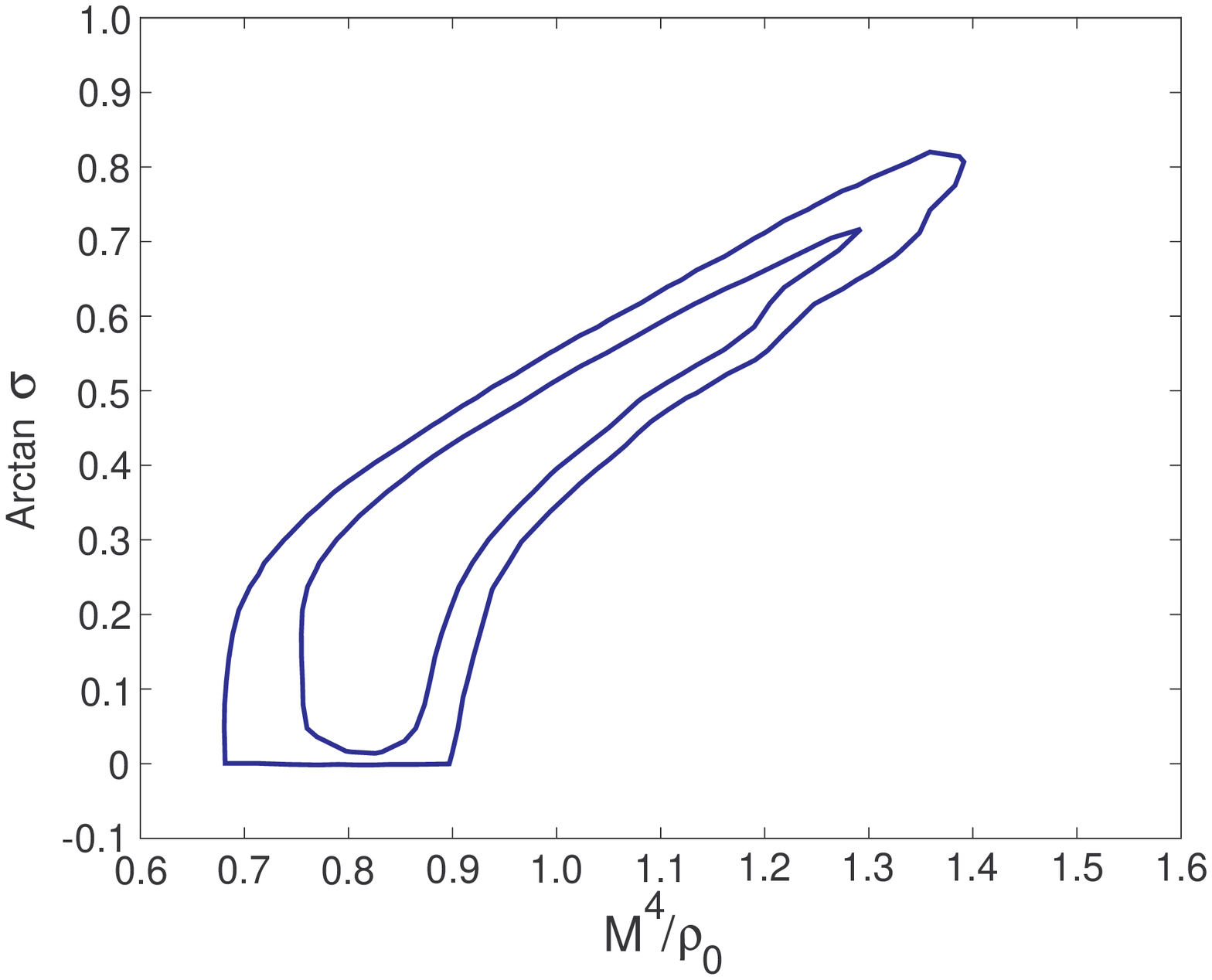}
\caption{Constant confidence contours (68\% and 95\%) in the
($M^4/\rho_0,\arctan \sigma$) plane allowed by SNeIa (left panel)
and X-ray galaxy clusters (right panel).} \label{snecluster}
\end{figure}

\section{Observational Constraints}
The zeroth order quantities (such as the luminosity and angular
diameter distances), depend only on integrals of the Hubble
parameter. Therefore, they are not very sensitive to local features
of the function $\rho\left(a\right)$. In particular, they should not
depend on the specific form of the transition from $p=0\,\,\,$to
$p=-M^{4}$. For example, for small values of $\sigma$, the
observational data (from SNIa, for instance) constrains only $M^{4}$
and not $\sigma$ nor $\beta$. Thus we expect the background
observational constraints to be highly degenerate for small $\sigma$
($\sigma\lesssim0.1$). Further, as will be shown, first order tests,
such as cosmic microwave background fluctuations or large-scale
structure data, constrain the value of $\sigma$ to be very small
($\sigma\ll1$). Therefore, a real step function is a good model
independent approximation for the background evolution in the type
of quartessence we are dealing with in this paper.
\begin{figure}
 \includegraphics[height= 6.5 cm,width=8.5cm]{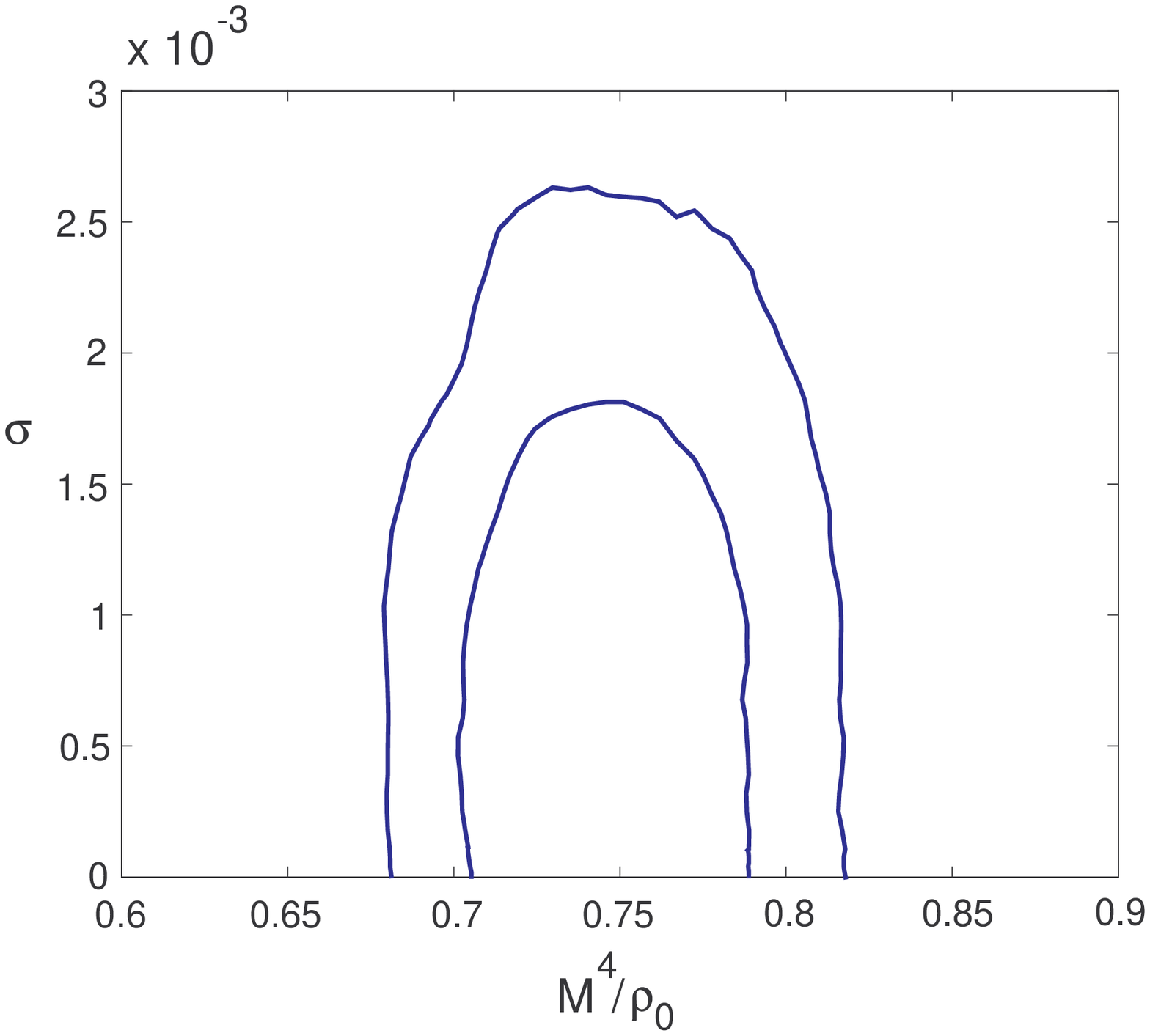} \includegraphics[height= 6.5 cm,width=8.5cm]{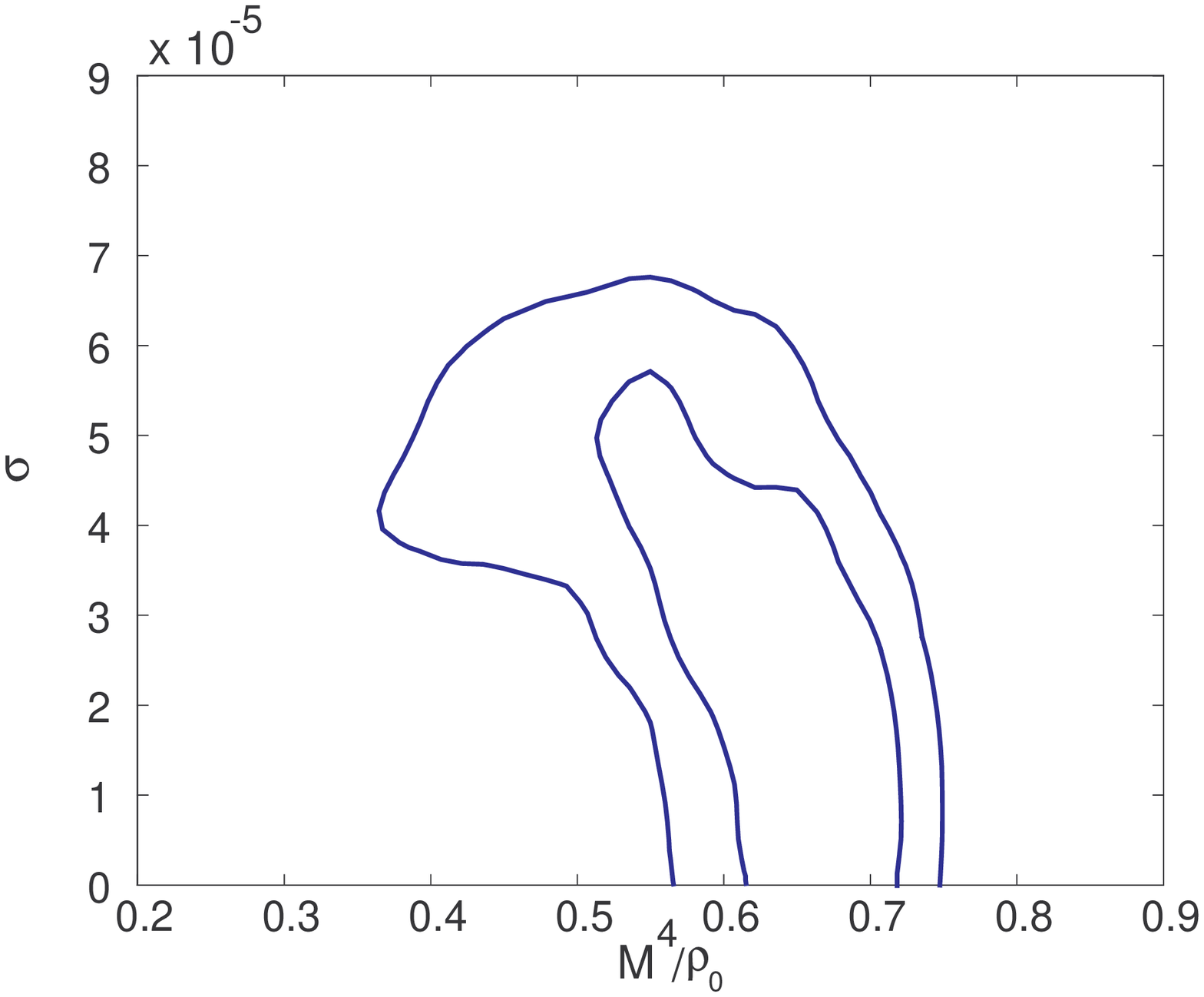}\\
\caption{Constant confidence contours (68\% and 95\%) in the
($M^4/\rho_0,\sigma$) plane allowed by CMB (WMAP) (left panel) and
matter power spectrum (SDSS) \cite{SDSS} (right panel).}
\label{1order}
\end{figure}

In the following we derive constraints on the parameters $\sigma$
and $M^{4}/\rho_{0}$ from four data sets: SNIa, X-ray cluster gas
mass fraction, galaxy power spectrum and CMB fluctuations. Here,
$\rho_{0}$ is the present value of the quartessence energy density.
For the sake of simplicity, in our computations we fixed the
parameter $\beta$ to the intermediary value $\beta=2$. We remark
that, for small values of $\sigma$, $-M^{4}/\rho_{0}\simeq w_0$,
where $w_0$ is the present equation of state parameter. It is worth
pointing out that $w_0$ should not be compared to the usual dark
energy EOS $w_{DE}$ but with $w_{\rm eff}\equiv
w_{tot}\Omega_{tot}$. In a flat Universe and neglecting the small
amount of baryons $w_{\rm eff}\equiv w_{DE}\Omega_{DE}$. Values
around $M^{4}/\rho_{0}\approx 0.7$ are therefore to be expected.

In our SNIa analysis we use the ``gold'' data set of Riess
 \textit{et al.} \cite{riess04}. To determine the likelihood of the
parameters we follow the same procedure described in
\cite{amendola05} assuming flat priors when marginalizing over the
baryon density parameter $\Omega_{b0}h^{2}$ and Hubble parameter
$h$. For the galaxy cluster analysis, we use the \emph{Chandra}
measurements of the X-ray gas mass fraction data from Allen
\textit{et al.} \cite{allen04}. Again, we follow the same procedure
described in \cite{amendola05} to determine confidence region of the
parameters of the model. We first marginalize analytically over the
bias $b$, using a Gaussian prior with $b=0.824\pm0.089$ and then, as
in the SNIa analysis, we marginalize over $\Omega_{b0}h^{2}$ and $h$
assuming flat priors. In figure \ref{snecluster} we show constant
68\% and 95\% confidence levels contours on the parameters
$M^{4}/\rho_{0}$ and $\sigma$ for SNIa and X-ray galaxy clusters.
From the figure it is clear that, as expected, background tests
impose only weak constraints on the parameter $\sigma$.

In order to obtain constraints on $M^{4}/\rho_{0}$ and $\sigma$ from
CMB data \cite{wmap} we follow the procedure described in
\cite{amendola05}, fixing $T_{CMB}=2.726K$, $Y_{He}=0.24$ and
$N_{\nu}=3.04$, and marginalizing over the other parameters, namely,
$\Omega_{b0}h^{2}$, $h$, the spectral index $n_{s}$, the optical
depth $\tau$ and the normalization $N$. In figure \ref{1order} (left
panel) we show the confidence region on the parameters for CMB. Note
that $\sigma$ plays a decisive role in the evolution of
perturbations; now the data constrain this parameter to be
$\sigma\lesssim3\times10^{-3}$.

We next consider the matter power spectrum, comparing the baryon
spectrum with data from SDSS \cite{SDSS}. To compute the likelihood,
we used a version of the code provided by M. Tegmark
\cite{sloancode}, cutting at $k=0.20\; h$Mpc$^{-1}$ ($19$ bands) and
marginalizing over $\Omega_{b0}h^{2}$, $h$, $n_{s}$  and the
amplitude. In figure \ref{1order} (right panel) we show the 68\% and
95\% confidence levels on $\sigma$ and $M^{4}/\rho_{0}$ from the
SDSS power spectrum. This is the most restrictive test we have
considered in this work, implying that
$\sigma\lesssim7\times10^{-5}$.

\begin{figure}
 \includegraphics[height= 6.5 cm,width=8.5cm]{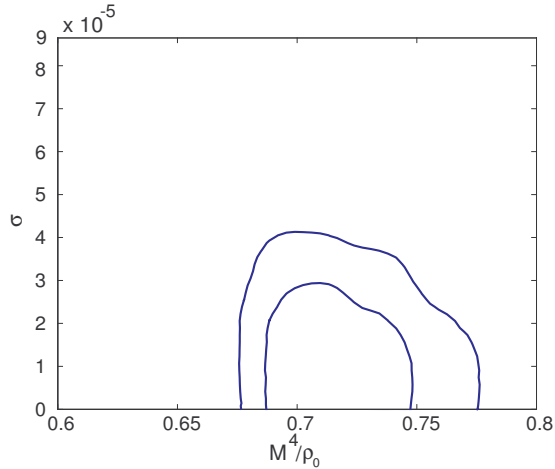}
\caption{68\% and 95\% in the ($M^4/\rho_0,\sigma$) plane for the
combined analisys SNIa + galaxy clusters + matter power spectrum +
CMB.} \label{comb}
\end{figure}

In figure \ref{comb} we display the constant (68\% and 95\%)
contours for the combined analysis SNIa + X-ray galaxy clusters +
matter power spectrum + CMB data. Our final result ($95\%$) is
$0.68\lesssim M^{4}/\rho_{0}\lesssim0.78$ and
$0<\sigma\lesssim4\times10^{-5}$. It is straightforward to show that
the transition redshift from a pressureless epoch to a constant
negative pressure period is given by
$z_t\simeq[(M^{4}/\rho_{0})(1-\sigma)/((1-M^{4}/\rho_{0})\sigma)]^{1/3}$.
Therefore, assuming $M^{4}/\rho_{0}\sim0.7$ and since
$\sigma\lesssim4\times10^{-5}$ the transition from $p=0$ to
$p=-M^{4}$ would have occurred at $z_{t}\gtrsim38$.

\section{Conclusion}
In this work we presented a new adiabatic quartessence model
characterized by a change of concavity in the EOS. We obtained the
constraints on the model parameters from SNIa, X-ray gas mass
fraction in galaxy clusters, CMB and matter power spectra and showed
that the model is viable if $\sigma\lesssim4\times10^{-5}$. The
redshift of the transition from the regime $p\simeq0$ to $p\simeq
const.<0$ is, therefore, $z_t\gtrsim38$. On the other hand, the
inclusion of matter power spectrum data for smaller scales
($k\gtrsim0.2$ h Mpc$^{-1}$) could impose stronger constraints upon
$\sigma$ pushing the minimum redshift of the transition to higher
values. We checked that this is, in fact, the case by considering
data from the matter power spectrum from the Lyman-alpha forest
\cite{lya}. However, since there are still systematic uncertainties
in this data we did not include them in our analysis. Although
differences between quartessence models and $\Lambda$CDM may exist
in the nonlinear regime \cite{dimitrius}, the results of the present
work, in combination with the results of \cite{sandvik02} and
\cite{amendola05}, indicate that, at zero and first orders, any
(convex or not) successful adiabatic quartessence model cannot be
observationally distinguished from $\Lambda$CDM.

\acknowledgments{We thank Roberto Colistete and Miguel Quartin for
useful discussions. The CMB computations have been performed at
CINECA (Italy) under the agreement INAF@CINECA. We thank the staff
for support. RRRR and IW are partially supported by the Brazilian
research agencies CAPES and CNPq, respectively. LA thanks the Gunma
Nat. Coll. of Techn. (Japan) for hospitality during the later stages
of this work and JSPS for financial support.}


\begin{thebibliography}{10}
\bibitem{makler03}M. Makler, S.Q. Oliveira, and I. Waga, Phys. Lett. B \textbf{555},
1, 0209486 (2003).
\bibitem{kamenshchik01}A. Kamenshchik, U. Moschella, and V. Pasquier, Phys. Lett. B \textbf{511},
265 (2001); M. Makler, \textit{Gravitational Dynamics of Structure
Formation in the Universe}, PhD Thesis, Brazilian Center for
Research in Physics (2001); N. Bili\'{c}, G.B. Tupper, and R.D.
Viollier, Phys. Lett. B \textbf{535}, 17 (2002); M.C. Bento, O.
Bertolami, and A.A. Sen, Phys. Rev. D \textbf{66}, 043507 (2002).
\bibitem{makler03b}M. Makler, S.Q. Oliveira, and I. Waga, Phys. Rev. D \textbf{68},
123521 (2003); A. Dev, D. Jain, and J.S. Alcaniz, Astron. Astrophys.
\textbf{417}, 847 (2004); Z.-H. Zhu, Astron. Astrophys.
\textbf{423}, 412 (2004); R. Colistete Jr. and J.C. Fabris, Class.
Quant. Grav. \textbf{22}, 2813 (2005); M.C. Bento, O. Bertolami,
N.M.C. Santos, and A.A. Sen, Phys. Rev. D \textbf{71}, 063501
(2005).
\bibitem{carturan03}D. Carturan and F. Finelli, Phys. Rev. D \textbf{68}, 103501 (2003);
L. Amendola, F. Finelli, C. Burigana and D. Carturan, JCAP
\textbf{07}, 005 (2003).
\bibitem{sandvik02}H.B. Sandvik, M. Tegmark, M. Zaldarriaga, and I. Waga, Phys. Rev. D\textbf{69},
123524 (2004).
\bibitem{reis03b}R.R.R. Reis, I. Waga, M.O. Calv\~{a}o, and S.E. Jor\'{a}s, Phys.
Rev. D \textbf{68}, 061302(R) (2003).
\bibitem{amendola05}L. Amendola, I. Waga, and F. Finelli, JCAP \textbf{11}, 009 (2005).
\bibitem{reis05}R.R.R. Reis, M. Makler, and I. Waga, Class. Quantum Grav. \textbf{22},
353 (2005).
\bibitem{bento}M.C. Bento, O. Bertolami and A.A. Sen, Phys. Rev. D
\textbf{70}, 083519 (2004).
\bibitem{bilic}N. Bilic, G.B. Tupper and R.D. Viollier,
hep-th/0504082.
\bibitem{riess04}A.G. Riess  \textit{et al.}, Astrophys. J. \textbf{607}, 665 (2004).
\bibitem{allen04}S.W. Allen \textit{et al.}, Monthly Notices of the Royal Astron.
Society \textbf{353}, 457 (2004).
\bibitem{wmap}G. Hinshaw \textit{et al.} (the WMAP collaboration),
Astrophys. J. Suppl. \textbf{148}, 135 (2003); L. Verde \textit{et
al.} (the WMAP collaboration), Astrophys. J. Suppl. \textbf{148},
195 (2003).
\bibitem{SDSS}M. Tegmark, \textit{et al.} (the SDSS collaboration), Phys. Rev. D
\textbf{69}, 103501 (2004); M. Tegmark, \textit{et al.} (the SDSS
collaboration), Astrophys. J. \textbf{606}, 702 (2004).
\bibitem{sloancode}http://space.mit.edu/home/tegmark/sdss.html
\bibitem{dimitrius}D. Giannakis and W. Hu, Phys. Rev. D \textbf{72}, 063502 (2005).
\bibitem{lya}R.A.C. Croft \textit{et al.}, Astrophys. J. \textbf{520}, 1 (1999);
N.Y. Gnedin and A.J.S. Hamilton, Monthly Notices of the Royal
Astron. Society \textbf{334}, 107 (2002).
\end{thebibliography}
\end{document}